\documentclass[prd,twocolumn,floatfix]{revtex4}
\usepackage[dvips]{graphics}
\usepackage{amsmath,amsfonts}
\usepackage{psfig}

\begin{document}

\newcommand{\bra}{\langle}
\newcommand{\ket}{\rangle}
\newcommand{\beq}{\begin{equation}}
\newcommand{\eeq}{\end{equation}}
\newcommand{\be}{\begin{equation}}
\newcommand{\ee}{\end{equation}}
\newcommand{\bea}{\begin{eqnarray}}
\newcommand {\eea}{\end{eqnarray}}
\def\bea{\begin{eqnarray}}
\def\eea{\end{eqnarray}}
\def\tr{{\rm tr}\,}
\def\erf{{\rm erf}\,}
\def\sgn{{\rm sgn}\,}
\def\href#1#2{#2}

\title{ Relaxing to Three Dimensions}

\author{Andreas Karch}
\affiliation{ Department of Physics,
University of Washington,
Seattle, WA 98195, USA}
\author{Lisa Randall}
\affiliation{ Department of Physics,
Harvard University,
Cambridge, MA, 02138, USA}

\begin{abstract}
We propose a new selection principle for distinguishing among possible vacua that we
call the ``relaxation principle.'' The idea is that the universe will
naturally select among possible vacua through its cosmological
evolution, and the configuration with the biggest
filling fraction is the likeliest.
We apply this idea to the question of the number of dimensions
of space. We show that under conventional (but higher-dimensional) FRW
evolution, a universe filled with equal numbers of branes and antibranes
will naturally come to be dominated by 3-branes and 7-branes.
We show why this might explain the number of dimensions that are
experienced in our visible universe.

\end{abstract}
\today
\maketitle

\noindent{\bf Introduction:}
No fundamental physical principle singles out three dimensions. Yet
three dimensions of space clearly has a special status. The obvious
question is why.

There  have been several  attempts to address this question,
particularly in the context of string theory. Probably the idea that
has received the most attention is the Brandenberger-Vafa suggestion \cite{bv}.
They argue that four spacetime dimensions are singled out because
the world sheet of a string occupies two dimensions and in four dimensions two two-dimensional worldsheets
will intersect. Their argument is
that unless the worldsheets can intersect, nothing  dilutes the
winding modes of strings. And if nothing dilutes the winding modes, the
initially very small dimensions would never
grow large. However, if there are four (or fewer) spacetime dimensions, strings can
shed their windings so that the dimensions can grow.

This is an interesting scenario, but has several questionable
aspects. One is that it critically relies on poorly-understood
dynamics at the Planck scale. Another is that it requires a
resolution of the moduli problem, and, more importantly, relies on
simple toroidal compactification. But the most problematic feature
of this solution is that it relies on strings being the sole
important objects in string theory, whereas we now know that
branes also play a critical role.
Some work has been done on addressing this concern, see in
particular \cite{steph}.

Another more recent suggestion by \cite{intersect}
was also made in the context
of string theory with compact dimensions.  They suggest another
reason that four spacetime
dimensions are special. They point out that $4+4<10$
and that this not true for any larger
integer. Their argument was that the worldvolume of 3-branes would NOT intersect whereas the worldvolume of any
larger branes would. Their argument was that larger branes can therefore unwind, whereas 3-branes would
survive.
Their suggestion is very interesting but has some technical
problems. In particular, the unwinding mechanism they suggest is
flawed--the branes would generally merge to form new (p,q)-type
branes.
The same numerical fact has also been used in \cite{quevedo},
where the universe is created via a sequence of brane/antibrane
annihilations. The observation that $4+4<10$ is of course robust. Here we
will use this numerical fact in an entirely different context,
with NO compact dimensions.

Branes and localization have opened
up an entirely new way of approaching higher-dimensional cosmology,
since it is possible that both gravitational
and nongravitational forces can be localized on 3-branes.  Yet very
few cosmological studies have been done that exploit this idea.
We now know that the apparent number of dimensions isn't necessarily the same as the
number of dimensions that actually exists,
so it is reasonable to think the initial evolution of the universe might be  higher-dimensional.

We invoke higher-dimensional cosmology in conjunction with what we call the ``relaxation principle.'' The idea  is
that there are many conceivable vacua, as would be described, for
example, in a landscape scenario. But rather than invoking the
anthropic principle to chose among them, we suggest that the ones
that are favored are those   that  dominate in a consistent stable
cosmological scenario. One attempt in this
direction in the context of the cosmological constant was given in \cite{shinji}.

In the context of brane gas
dynamics, we assume that the branes with the biggest filling
fraction in the endpoint of the universe's cosmological evolution
are the most likely branes to be relevant to the state in which we
live.  String theory does indeed indicate there could be multiple vacua.
This makes it especially
worthwhile to investigate  alternative
selection principles. It is possible that dynamics, rather than static
criteria, determines the vacuum in which we live.

In this letter we show that under some very general and
plausible assumptions about brane dynamics in ten  dimensions,
 branes with other numbers of dimensions will be diluted relative to the
3-branes and 7-branes.
In our approach to cosmological evolution, we assume the number of
dimensions
is fixed but that the number of dimensions we see is determined by
brane dynamics.
We make
a general ansatz that the universe is initially filled with branes of all
sizes and that the  stress-energy on the branes dominates the cosmological evolution. We then
let the universe expand and look for a consistent solution. We find the
only stable evolution under reasonable general assumptions
is into a universe dominated by 3-branes and
7-branes. As in previous work, we will use the fact that 3-branes will be the largest objects whose
self-intersections will not significantly reduce their density (since $4+4<10$).

Recent work on cosmic strings  proceeds along similar lines to the
analysis we present below. That work assumes three large
dimensions whereas ours assumes  nine. Our
broader framework might be more appropriate in string theory now that we
know that branes play a critical role. However, strings  could
play an important role along the lines of \cite{pol} once the
universe has settled into an effective four-spacetime-dimensional
theory--that is, once all the other branes have diluted and
gravity has localized on the brane intersection.

\noindent{\bf FRW Universe in higher dimensions:}
Starting with the $n+1$ dimensional Robertson Walker metric
\beq ds^2 = - dt^2 + a^2(t) d \Sigma_k^2 \eeq
with $n$ dimensional maximally symmetric spatial geometry $\Sigma_{k=-1,0,1}$,
the Friedmann equations dimensions read:
\beq
\label{frw} H^2 = -\frac{k}{a^2} + \frac{16 \pi G_N}{n (n-1)} \rho \eeq
Energy conservation demands
\beq \frac{\dot{\rho}}{\rho} = - n (1+ w) \frac{\dot{a}}{a} \eeq
where $w$, as usual, determines the equation of state of the
matter system, $p = w \rho$. If the right hand side of
(\ref{frw})
is dominated by a component of given $w$ we can solve for
the time dependence of $a$ and $\rho$ in that era:
\beq \label{aoft}
 \rho \sim  a^{-n (1+w)}  \rightarrow   t \sim a^{ \frac{n}{2} (w+1) } \eeq

A few typical values of $w$ are $w=0$ for pressureless dust,
$w=1/n$ for radiation with traceless stress tensor, a cosmological
constant has $p=-\rho$ and hence $w=-1$ in any $n$. Note that
$w_{crit.}=-\frac{n-2}{n}$ sets the borderline between
acceleration and deceleration. More interesting for us are the $w$
values of networks of topological defects, also known as brane
gases. A string has $\rho=-p$ like a cosmological constant, but
only one component of $p$ is non-zero, so with strings in random
directions in $n=3$ spatial dimensions the average $p$ is going to
be $-\frac{1}{3} \rho$. In the same way one can argue that a
$d$-brane with a $d+1$ dimensional worldvolume in $n$ spatial
dimensions has $w=-\frac{d}{n}$. For a non-interacting gas of
$d$-branes according to (\ref{aoft}) the energy density goes as
$\rho^{ni}_d \sim a^{d-n}$; the volume of the brane goes as $a^d$,
but the volume of space goes like $a^n$ so the energy density goes
as $a^{d-n}$.

However, if the branes can self-intersect we expect them to decay
and hence to dilute much faster. In the familiar case of $d=1$
strings in $n=3$ spatial dimensions, the decay mechanism is that
strings that intersect spawn loops of closed string which then
decay by emitting gravity waves, for a review see \cite{pol}. To
see how the energy density of such a self-intersecting brane
network behaves as a function of time one can follow the same very
general line of logic that is usually applied for cosmic strings:
assuming that the decay processes happen at the maximum efficiency
allowed by causality, the network at any time looks the same when
viewed at the horizon scale $t$. This is often referred to as the
scaling solution. The total length of string within a horizon
volume is hence some number times $t$.
Similarly the total area of a two-dimensional membrane is some
number times $t^2$ and the total volume of $d$-brane some number
times $t^d$. The horizon volume is just given by $t^n$, hence
$\rho_d^i \sim t^{d-n}$. In order to determine which type of
defect is going to dominate the energy density, we need to know
which defects will interact and which not. We assume that the
generic situation is that any defects that can find each other
will interact and trigger some decay mechanisms that will work
with an efficiency only limited by causality. The efficiency with
which branes and antibranes that find each other annihilate is a
critical question that we would like to see checked in future
work.

The question of whether branes can generically find each other
depends only on their dimensionality. In $n+1$ dimensions generic
(that is no parallel directions) $d$-branes with $2d \geq n$
intersect at all times over a $2d-n$-brane. For $2d=n-1$ (e.g.
strings in $n=3$), they intersect over a (-1) brane, that is they
collide at an instant. $d$-branes with $2d \leq n-2$ on the other
hand generically do not find each other. For $n=3$ we reproduce
the well known results that monopoles generically do not find each
other, while cosmic strings and domain walls do. For the case
$n=9$, which is the critical dimension of superstring theory, we
see that branes with $d \leq 3$ will not find each other, while
generic branes with $d \geq 4$ at least intersect at an instant in
time. This simple counting was already proposed in
\cite{intersect} as a possible mechanism to select a 4-dimensional
world, alas in the context of torus compactifications where the
branes have conserved winding charges and hence generically cannot
annihilate.

We also note an additional feature of branes that affects their
density. Brane-antibrane pairs generically contain a tachyon. In
the case of 9-branes, that tells us that 9-branes will generically
annihilate since they overlap completely at all times. So 9-branes
will not have energy and number density that scales as above, but
will have zero density for all times.

The case of 8-branes is trickier. It might be that 8-branes never
exist since it is possible that only even or odd dimensional
branes exist. This is true for D-branes in type II string theory.
It would also be true of the tachyon is a complex scalar and lower
dimensional branes are produced as defects in
higher-dimensional-brane annihilation. This is analogous to the
reasoning in \cite{pol}, for example, where it was argued that
strings but not domain walls might be present in the cosmos
(domain walls give exactly the same problem in four-dimensions
that 8-branes give us here). Furthermore, another feature of
string theory might affect 8-branes. In flat space, a static
 8-brane/anti-8-brane configuration cannot have branes that are
very far from each other or else spacetime breaks down. If this is
also true for dynamical systems in arbitrary backgrounds, it would
mean that 8-branes are also expected to have number density less
than $t^{8-n}$.

We now turn to lower-dimensional branes. This is where the significance of
three dimensions becomes apparent.
Were there no brane intersections, the higher-dimensional branes
would dilute much more slowly. However, the intersections change
the dilution so that lower-dimensional and higher-dimensional
branes can compete. By assuming that if the branes intersect, they will
annihilate and dissipate their energy at a rate determined by causality,
we can compare the densities of the branes of different dimensionalities.

 For $d\le 3$, the branes dilute as
$a^{d-n}$, whereas for higher $d$, they scale as $t^{d-n}$. Among
the branes with $d \le 3$, it is clear that the 3-branes will
dilute the least. To resolve the competition between $d \le 3$ and
higher dimensional branes, we need to know the relation between
$a$ and $t$, or alternatively the $w$ of the dominant fluid
component. For any $w<w_{crit} = -\frac{n-2}{n}$, (\ref{aoft})
tells us that $t$ indeed grows slower than $a$. One option is to
take the 3-brane energy density to dominate, $w=-\frac{1}{3}$, and
to see if this is a self consistent solution. The resulting
time-dependence of the scale factor is $a \sim t^{\frac{1}{3}}$,
so it is clear that only 7-branes and 8-branes can compete with
3-branes. In fact, 7-branes dilute at the same rate and 8-branes
at a slower rate.  For the reasons given earlier, it is reasonable
to assume that 8-branes don't exist.

Really since we have assumed constant and
equal energy densities, the energy density represents filling
fraction in the universe. That is, since 3-branes and 7-branes
dominate the energy density and hence the evolution of the
ten-dimensional universe, they also dominate the filling-fraction
of the universe. Therefore, a reasonable hypothesis would be that
they are the most likely place for our universe to reside.

The result that 3-branes will eventually dominate is robust even
if we start with a different fluid component driving the
expansion. 3-branes have $w=-\frac{1}{3}$. If the universe is
dominated by $w>-\frac{1}{3}$ (like $w=0$ dust or $w=\frac{1}{9}$
radiation) all that happens is that the 7-branes can no longer
keep up.  For
$w>\frac{1}{3}$ 3-branes always dominate (assuming 9-branes annihilate).
Note that this in particular implies that brane fluctuations and
bending get washed out. The bending of the branes can be thought
of as a massless scalar living on the worldvolume of the branes.
Its gradient energy acts like a $w=1$ fluid and will hence be
subdominant. So independent of which component dominates early on,
a universe with 3-branes and 7-branes will be the only fixed point
of the evolution.  For the rest of the
paper we restrict our attention to this universe filled with
3-branes and 7-branes.


\noindent {\bf Applications:}
In the previous section we have shown that the dynamics of brane
networks will lead to a 10d universe dominated by 3- and 7-branes.
This calculation relied only on some simple dynamical assumptions 
about an expanding ten-dimensional universe, which should be explored further in the future. First was
 a homogenous and isotropic 10d universe, which is really
a statement about the initial conditions we assume. The precise 
evolution we used depends on initial conditions and 
could be accounted for by a stage of ten-dimensional inflation.  We also
assumed that any brane and antibrane that intersects will annihilate, which also merits
further investigation in the future. Finally, we need
the dilaton to be stabilized. It is known that in the presence of
an evolving dilaton the analysis of brane decay will be
significantly altered \cite{dilaton1,dilaton2}.

We now want to consider the physical implications of this result.
First recall that 3-branes and 7-branes are very interesting for
several reasons. Four-dimensional gauge theories with matter naturally exist with
3-branes and 7-branes, including the supersymmetric standard model
\cite{sm}. The universe is automatically full of just the right
ingredients to give the forces of nature.
The large number of D3 branes could also provide a natural
realization of AdS/CFT. This AdS space could play a critical role
in localizing gravity \footnote{We thank Bobby Acharya for
suggesting an interesting model of this sort.}
D3-branes and D7-branes
are also important for recent ideas about string theory models of
inflation \cite{kallosh, kklt,kklmmt}.
Recent  string-theoretic inflation models are based on the
presence of 3-branes and 7-branes \cite{kallosh,kklt,kklmmt}.
As we will see shortly,
this scenario could naturally give rise to this type of inflation.

The major issue we have  left to address is the origin
of four-dimensional gravity. We now suggest several reasons that four-dimensional gravity
can exist,
leaving the details for a forthcoming publication.
Perhaps the least interesting possibility is that dimensions are
in fact compact. One way our analysis could apply in that case is
if nine spatial dimensions expand to a large size in the early
evolution of the universe, and only afterwards stabilize at a
small size.
%
Or the universe gets effectively compactified without moduli
fields due to the presence of the 7-branes, which can effectively
compactify dimensions as in Ref. \cite{vafa}. 7-branes are
codimension 2 branes in 10d and as such they change the metric not
just in their local neighborhood, but also affect the global
geometry. For the generic 7-brane solution, for example the D7
brane of IIB string theory or a fat codimension 2-brane, as we
will analyze below in the context of localized gravity, the metric
naively becomes singular a finite distance away from the 7-brane.
In order to see if the singularity can get resolved, one needs
global information about the spacetime behind the singularity. In
the case of D7 branes global solutions are known \cite{vafa} in
which for a finite number of 7-brane the space transverse to the
7-branes becomes compactified into a space that serves as the base
of an elliptically fibered Calabi Yau. The geometry of the full
Calabi-Yau encodes in addition the behavior of the scalar fields,
the dilaton and the axion. The dimensionality of the compact space
depends on the orientation of the 7-branes. If all 7-branes are
parallel, the transverse 2d space becomes an $S^2$ base of an
elliptically fibered K3. If they intersect over a
$10-2k$-dimensional spacetime, the transverse $2k$-dimensional
space has to be the base of a Calabi-Yau $(k+1)$-fold.
This structure is governed by supersymmetry. It requires knowledge
beyond the local form of the 7-brane metric. It is conceivable
that in non-supersymmetric settings a globally consistent solution
allows the local 7-brane geometry to be pasted into a compact
space times a 4d FRW part.

It is encouraging for this scenario that the recent string theory constructions of
\cite{kklt,kklmmt} yielding compactifications with all moduli
fixed and leading to quasi-realistic cosmologies rely on IIB
compactification with 7-branes together with a huge number of
3-branes or fluxes carrying 3-brane charge. In these scenarios one
starts out with the internal space being compact and only follows
the cosmological evolution in the effective 4d low energy theory.
One can easily imagine this being the outcome of a 10d
cosmological evolution that led 7- and 3-branes to dominate,
presumably also determining which  type of geometry is
preferred. If for some reason the 3-branes tend to clump, in this
case we can also get a scenario similar to that of Verlinde
\cite{verlinde}. The 3-branes pull an AdS$_5$ $\times$ $S^5$
throat out of the compact manifold, naturally stabilizing a large
hierarchy.

However, perhaps the most promising scenario exploits the notion of localized gravity, which
fits in naturally with our assumption that we can neglect
compactification in the universe's early evolution. A
configuration that is a natural candidate for four-dimensional
gravity is the intersection of three 7-branes where the
intersection has spacetime dimension four.  One can have a triple
7-brane intersection where the intersection has spacetime
dimension four. It was shown in \cite{gherghetta} that a single
codimension 2-brane can localize gravity in an embedding AdS bulk
space, so long as the brane satisfy a tuning conditions on its
energy density. Using the methods of Ref. \cite{nima}, one can
generalize this setup to the case of 3 intersecting codimension 2
branes in $AdS_{10}$. There is a normalizable graviton and three-dimensional gravity localized
at the interersection.
 
This shows one can consistently get three-dimensional gravity as well as forces localized
on three branes in this scenario. The
remaining challenge is to connect this configuration to the
analysis of the first part of this paper; that is, to find a
reason this configuration might be favored and to show why the forces should be coincident.

The problem is the
intersections have a smaller filling fraction than the 7-branes
themselves, or 5-brane intersections of two 7-branes for that
matter. We suggest two  reasons that the triple intersections might be
favored, both of which merit further investigation. The first
possibility is that the 3-branes play an important role in
creating the intersection. That is, if there is an attractive
force between 3-branes and 7-branes and they want to line up, this
can create the preferred geometry. This would have the added
feature of leading to gauge forces coincident with
four-dimensional gravity. The second possibility that is under
investigation is the possibility of four-dimensional "knots",
which act as locally localized gravitational regions.
 The idea is that in
addition to infinite 7-branes, there could be 7-brane loops  in
the directions orthogonal to the four overlapping dimensions on
which gravity is localized. In order to localize gravity,  the
loops would probably have to have size somewhat bigger than the
Planck length. They would therefore  require flux passing through
or on the branes themselves to stabilize them. The density of these finite-sized ocnfigurations
(in extra-dimensional space)
could be comparable to that of the fundamental 3-branes we considered earlier
since from a cosmological perspective these
7-brane balls look like 3-branes.   Another
advantage of these loop configurations might be that they could trap
3-branes in their interior, so that gravity and forces would be
coincident. In fact, this raises the possibility of naturally
explaining the hierarchy, since this would be the case if the
3-branes are near, but not precisely coincident, with the location
where gravity peaks.

We also wish to mention the interesting inflationary scenario this
model would suggest along the lines of
\cite{quevedo,braneinflation}.
Even if branes are attracted to the triple intersection, it is
likely that there is an initial excess of 3-branes or
anti-3-branes, giving rise to a large cosmological energy density.
Additional (anti-) 3 branes will aggregate at the intersection
later on and annihilate as in \cite{braneinflation}.
Because the extra dimensions are big (in the second scenario) or
infinite (in the first), one can naturally avoid some of the
constraints that frustrated the initial brane-inflation, which
doesn't last sufficiently long because the brane and antibrane
cannot start off at large enough separation.

%
%
%
%
%

\noindent{\large\bf Acknowledgments:} We would like to thank
A. Albrecht, P.
Creminelli, L. Fitzpatrick,
L. Motl, A. Nelson, A. Nicolis, E. Silverstein, C. Vafa, and L. Wang for
useful discussions, and S. Battersby for alerting us to the work of
\cite{intersect}. The work of AK was partially supported by the
DOE under contract DE-FGO3-96-ER40956.
The work of LR was supported in part by NSF Award PHY-0201124.

\bibliography{3brane}
\bibliographystyle{apsrev}
\end{document}